\documentclass[a4paper,10pt,english]{article}
\usepackage{nolta_cll,graphics,amsmath,amsfonts,array}
\usepackage{babel}
\usepackage[pdftex, unicode, bookmarks=true, bookmarksnumbered=true,
  pdfpagemode={UseOutlines}, plainpages=false, pdfpagelabels=true,
  colorlinks=true, linkcolor={black}, citecolor={black}, 
  urlcolor={black}]{hyperref}

\renewcommand\normalsize{\fontsize{9.7}{11.64}\selectfont}

\title{%
  Sample and Hold Errors in the Implementation of Chaotic Maps}

\author{%
  Sergio Callegari, Riccardo Rovatti\\
  \small DEIS, University of Bologna, Italy\\
  \small\texttt{(scallegari|rrovatti)@deis.unibo.it} 
  \thanks{This is a post-print version of a paper appeared in the
    Proceedings of the 1998 International Symposium on Nonlinear Theory
    and its Applications (NOLTA), vol.~1, pp. 199-202, Crans Montana (CH),
    Sept.~1998. To cite this paper, please use the published version data.}}

\date{}

\newcommand{\sh}{\operatorname{sh}}
\begin{document}

\begingroup
\def\footnotemark{}
\maketitle
\endgroup

\begin{abstract}
  \small Though considerable effort has recently been devoted to
  hardware realization of chaotic maps, the analysis generally
  neglects the influence of implementation inaccuracies. Here we
  investigate the consequences of S/H errors on Bernoulli shift, tent
  map and tailed tent map systems: an error model is proposed and
  implementations are characterized under its assumptions.
\end{abstract}

\section{Introduction}

\begin{figure}[ht]
  \setlength{\abovecaptionskip}{-0.2cm}
  \begin{center}
    \resizebox{0.8\linewidth}{!}{%
      \includegraphics{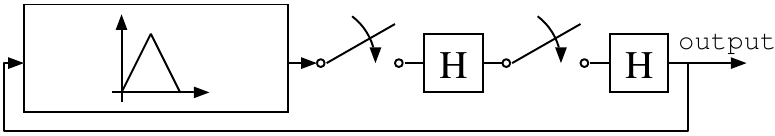}}
  \end{center}
  \caption{\label{fig:loop} \small Basic piecewise affine unidimensional
    chaotic loop. Two S/H stages are needed to hold the map input
    stable while the map circuit produces its output.}
\end{figure}

Silicon implementations of chaotic maps are analog discrete-time
systems exploiting sample and hold (S/H) stages to introduce the
necessary delay in the feedback loop (figure~\ref{fig:loop}).

Several efforts have recently been devoted to this area, identifying
current mode techniques as the most reliable and effective design
approach \cite{Delgado:EL-1993-25-2190, Langlois:NDES-1995,
Callegari:ISCAS-1997}. The many applications fields, including secure
communication \cite{Hasler:ISCAS-1994}, noise generation, stochastic
neural models \cite{Clarkson:WNNW93}, EMI reduction, etc., have
directed the attention mainly in obtaining interesting chaotic
behavious at the minimal hardware cost. On the contrary, the influence
of implementation inaccuracies has often been neglected, particularly
due to the difficulties inherent in relating the statistical
properties of chaotic systems to their implementation errors. Herein,
a sample and hold error model is developed and its correlation with
operating frequencies is suggested. Three maps characterized by a
uniform invariant probability density function (PDF) are then
considered --- the Bernoulli shift, the tent map, and the tailed tent
map--- and their implementation is characterized using this error
model. Particularly, influence of errors on the statistical properties
of the resulting signals is investigated, together with robustness
issues. As a conclusion, the a superior performance of the tailed tent map is
verified, and some design guidelines are drawn.  Focusing on S/H
errors is justified by the considerable overhead carried by most clock
feedthrough reduction techniques and by their general unscalability in
terms of the benefits one needs to achieve.

\section{Systems under investigation}

Without any loss of generality, we shall consider \emph{normalized}
maps, i.e.\@ maps whose invariant set (IS) is $[0,1]$. For a generic
map $M$ having IS $[x_h,x_l]$ a normalization function is defined as:
\begin{equation}
  \label{eq:normal}
  N(x)=(x-x_l)/(x_h-x_l)
\end{equation}
so that $M_n(x)=N(M(N^{-1}(x)))$ is the corresponding normalized map.

\begin{figure}[htb]
  \hspace*{\fill}
  \shortstack{%
    \resizebox{0.3\linewidth}{!}{\includegraphics{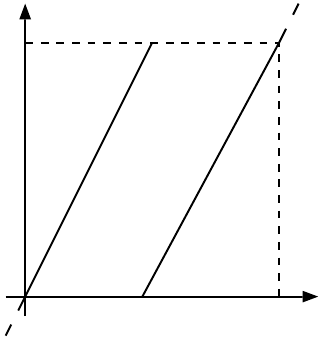}}\\[-0.15cm]
    \footnotesize(BS)} 
  \hfill
  \shortstack{%
    \resizebox{0.3\linewidth}{!}{\includegraphics{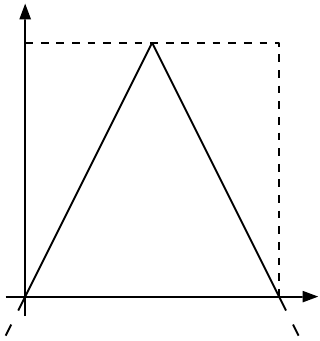}}\\[-0.15cm]
    \footnotesize(TM)} 
  \hfill
  \shortstack{%
    \resizebox{0.3\linewidth}{!}{\includegraphics{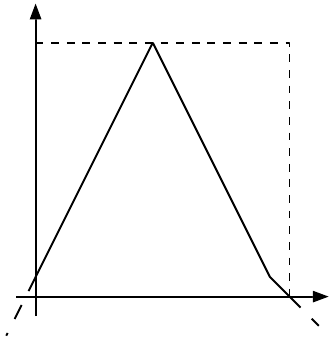}}\\[-0.15cm]
    \footnotesize(TTM)} 
  \hspace*{\fill}
  \caption{\label{fig:maps} \small Maps in the systems under
    investigation: Bernoulli shift, tent map and tailed tent
    map.}
\end{figure}

The systems under investigation are those based on the Bernoulli shift
(BS), the tent map (TM) and the tailed tent map (TTM)
\cite{Callegari:ISCAS-1997} (figure \ref{fig:maps}). All the systems
are characterized by a uniform PDF. The normalized maps are given by:
\begin{align}
  M_n(x)&=2x-\chi_{\{x>1/2\}}(x)\\
  M_n(x)&=1-2|x-1/2|\\
  M_n(x)&=1-2|x-\overline{t}/2|+\max(x-\overline{t},0)
\end{align}
respectively, where $\chi$ is a set characteristic function,
$\overline{t}=1-t$ and $t$ is a fraction which controls the
\emph{tail} size in the TTM.

In order to evaluate the robustness of implemented chaotic maps, one
must know how the map is defined out of its ideal IS
\cite{Callegari:ISCAS-1997}. For this purpose, we shall assume that
maps extend out of their definition set by linear extrapolation of
their edge branches (dashed lines in fig.~\ref{fig:maps}).

Focusing on a limited number of maps allows the use of extensive to
complement the achievable analyrical results.

\section{Modelling of sample and hold errors}

Since two sample and hold stages are necessary to implement the
feedback loop delay, we shall model both at once, up to the first
derivative, as in:
\begin{equation}
  \sh_n(x)=(1+\Delta\mu)x+\sigma_n-\frac{1}{2}\Delta\mu
\end{equation}
where $\Delta\mu$ is a slope error and $\sigma_n$ is a normalized
offset.
Note that, once again, generality is preserved by means of
normalization, so that the S/H model for a generic system would be  
\begin{equation}
  \sh(x)=N^{-1}\left((1+\Delta\mu)N(x)+\sigma_n-\frac{1}{2}\Delta\mu\right)
\end{equation}
where $N$ is the normalization function \eqref{eq:normal}. Since the
ideal S/H behaviour is $\sh(x)=x$, $\Delta\mu$ and $\sigma_n$ are
numeric indexes for the S/H error. Modelling up to the first
derivative represents a compromise between accuracy and the need to
keep error quantification simple and physically meaningful:
$\Delta\mu$ and $\sigma_n$ correspond to the common concepts of
\emph{signal dependent} and \emph{signal independent} S/H errors and
appear useful for the present analysis. However higher order
modelling would be necessary to accurately esteem errors on the
statistics of implemented systems, as it will appear further on.

Note that a viable way to consider S/H errors is thinking of the ideal
iteration of a perturbed map, which is a combination of the ideal map
and the S/H characteristic. Finally, notice that in many chaotic map
implementations the S/H stages represent the speed bottleneck due to a
accuracy/sample-latency tradeoff: for any given S/H circuit $\Delta\mu$
and $\sigma_n$ can be reduced only by increasing the memory
capacitance or by slowing the sample-to-hold commutation. Both actions
limit the cycle frequency.

\section{Characterization of implemented systems}

Characterization of the implemented chaotic maps will be given by
considering S/H\emph{s} as the sole error source and by looking at the
following features:

\vspace{-0.4cm}
\paragraph{Lack of robustness:}
shown by total loss of the system characteristic behaviour (loss of
chaoticity, acquisition of an IS which is not an interval).

\vspace{-0.5cm}
\paragraph{Alteration of the invariant set:} 
S/H errors usually alter the set in which a chaotic system produces
its samples.  If $[x_l,x_h]$ is the ideal IS and
$[\tilde{x}_l,\tilde{x}_h]$ is the IS due to S/H errors, the
normalized IS error:
\begin{equation}
  \epsilon_b=|N(\tilde{x}_h)-1|+|N(\tilde{x}_l)|
\end{equation}

In some applications, $\epsilon_b$ may not be critical.  Furthermore,
  it may be possible to track $\tilde{x}_h$ and $\tilde{x}_l$
  dynamically $\epsilon_b$ by linear rescaling.

\vspace{-0.5cm}
\paragraph{Alteration of the probability density function (PDF):}
expressed using an $L_1$ norm. If $\psi$ be the ideal PDF and
$\tilde{\psi}$ the real one, then the normalized PDF\emph{s} are:
\begin{align}
  \psi_n(x)&=(x_h-x_l)\,\psi(N^{-1}(x))\\
  \tilde{\psi}_n(x)&=(\tilde{x}_h-\tilde{x}_l)\,
    \tilde{\psi}(x\cdot(\tilde{x}_h-\tilde{x}_l)+\tilde{x}_l)
\end{align}
and the normalized PDF error is:
\begin{equation}
  \epsilon_\psi = \int_0^1|\psi_n(x)-\tilde{\psi}_n(x)|\,dx
\end{equation} 

Notice that this valuation of the PDF error implies linear
  compensation of $\epsilon_b$.

\vspace{-0.5cm}
\paragraph{Alteration of the cumulative probability density function (CDF):}
evaluated using an $L_1$ norm. If $\Psi$ is the ideal cumulative
probability density function (CDF) and $\tilde{\Psi}$ the real CDF,
then the normalized CDF\emph{s} are
\begin{align}
  \Psi_n&=\Psi(N^{-1}(x))\\
  \tilde{\Psi}_n&=
    \tilde{\Psi}(x\cdot(\tilde{x}_h-\tilde{x}_l)+\tilde{x}_l)
\end{align}
and the normalized CDF error is:
\begin{equation}
  \epsilon_\Psi = \int_0^1|\Psi_n(x)-\tilde{\Psi}_n(x)|\,dx
\end{equation}


\section{Results and analysis}

\begin{figure}[htb]
  \vspace{-0.5cm}
  \hspace*{\fill}
  \raisebox{-0.4cm}{\resizebox{0.5\linewidth}{!}{%
    \includegraphics{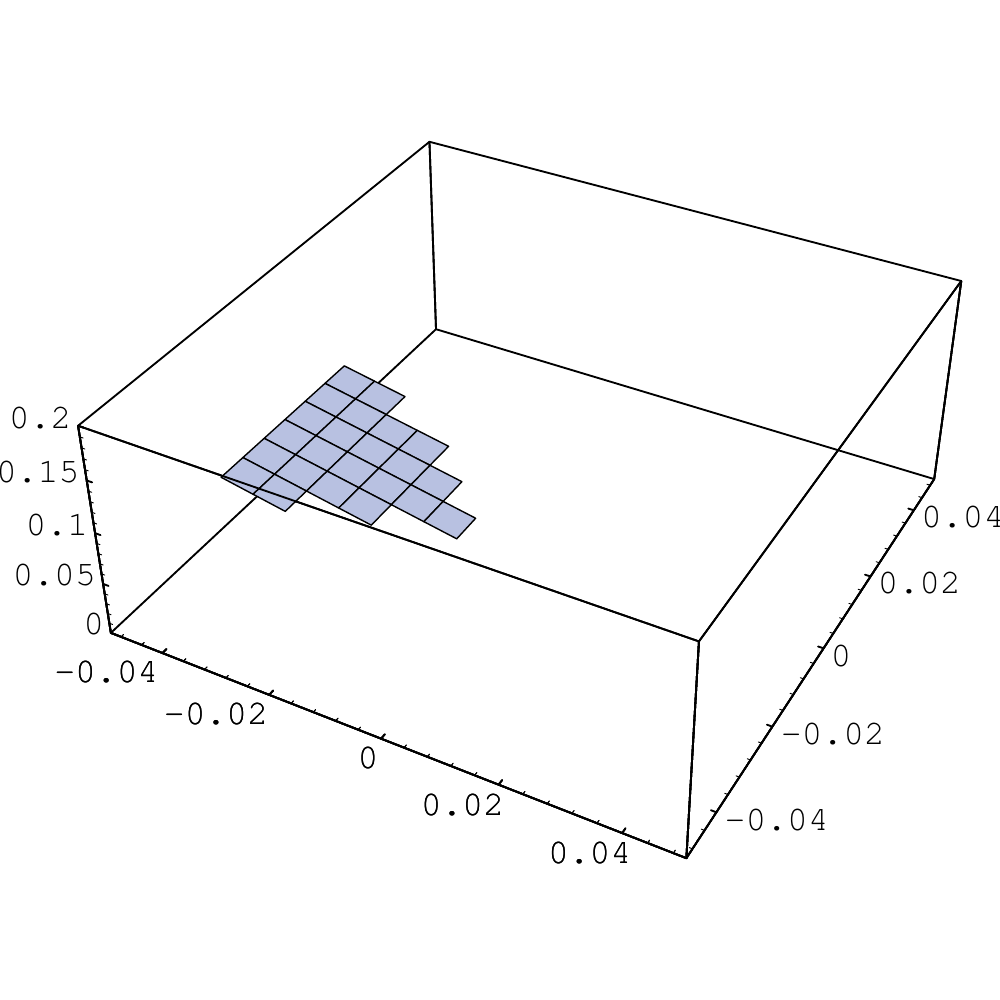}}}\
  \resizebox{0.42\linewidth}{!}{%
    \includegraphics{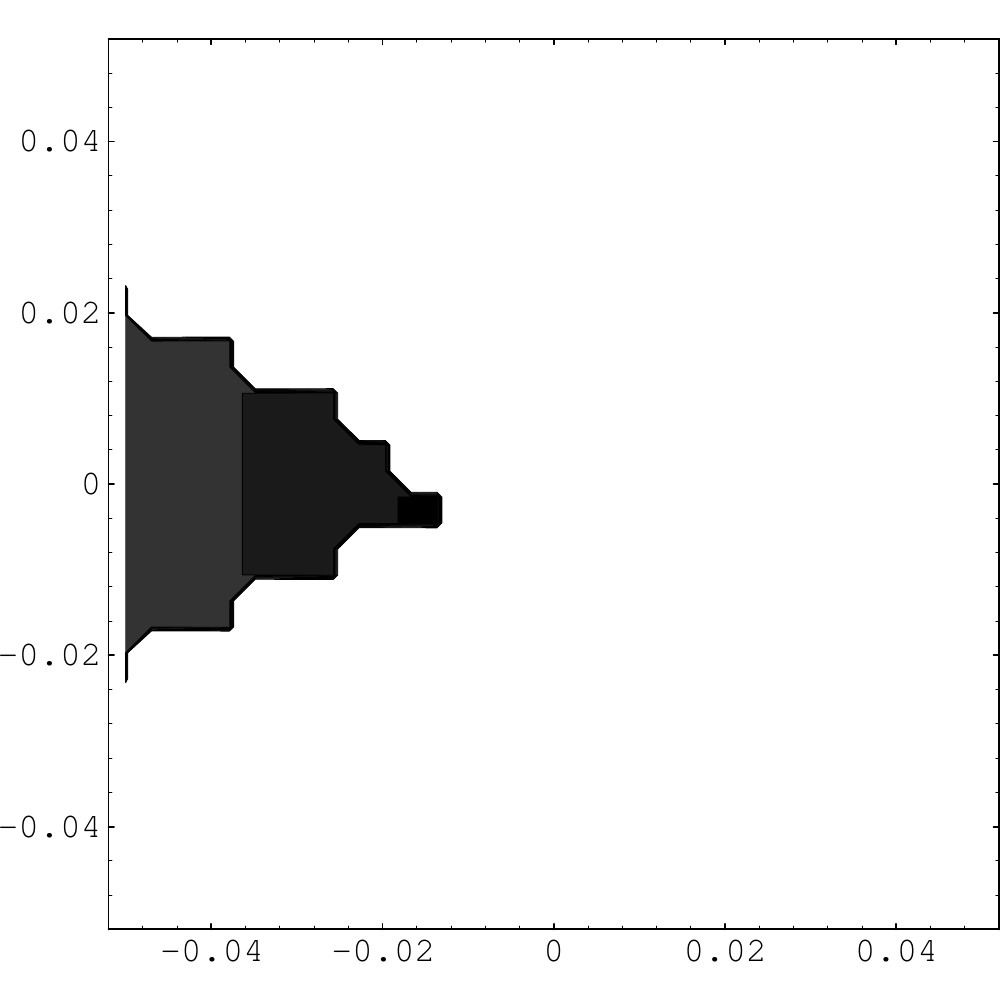}}\hspace*{\fill}\\[-0.8cm]
  \hspace*{\fill}
  \raisebox{-0.4cm}{\resizebox{0.5\linewidth}{!}{%
    \includegraphics{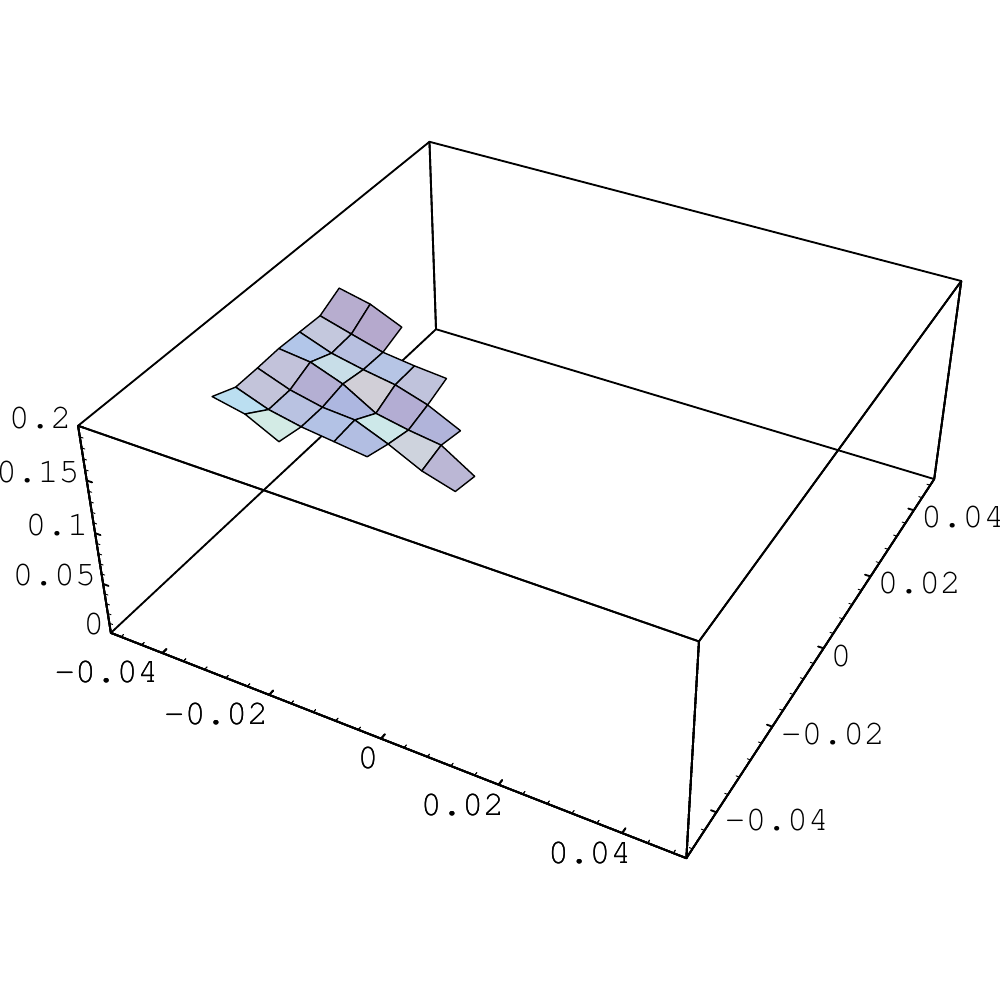}}}\
  \resizebox{0.42\linewidth}{!}{%
    \includegraphics{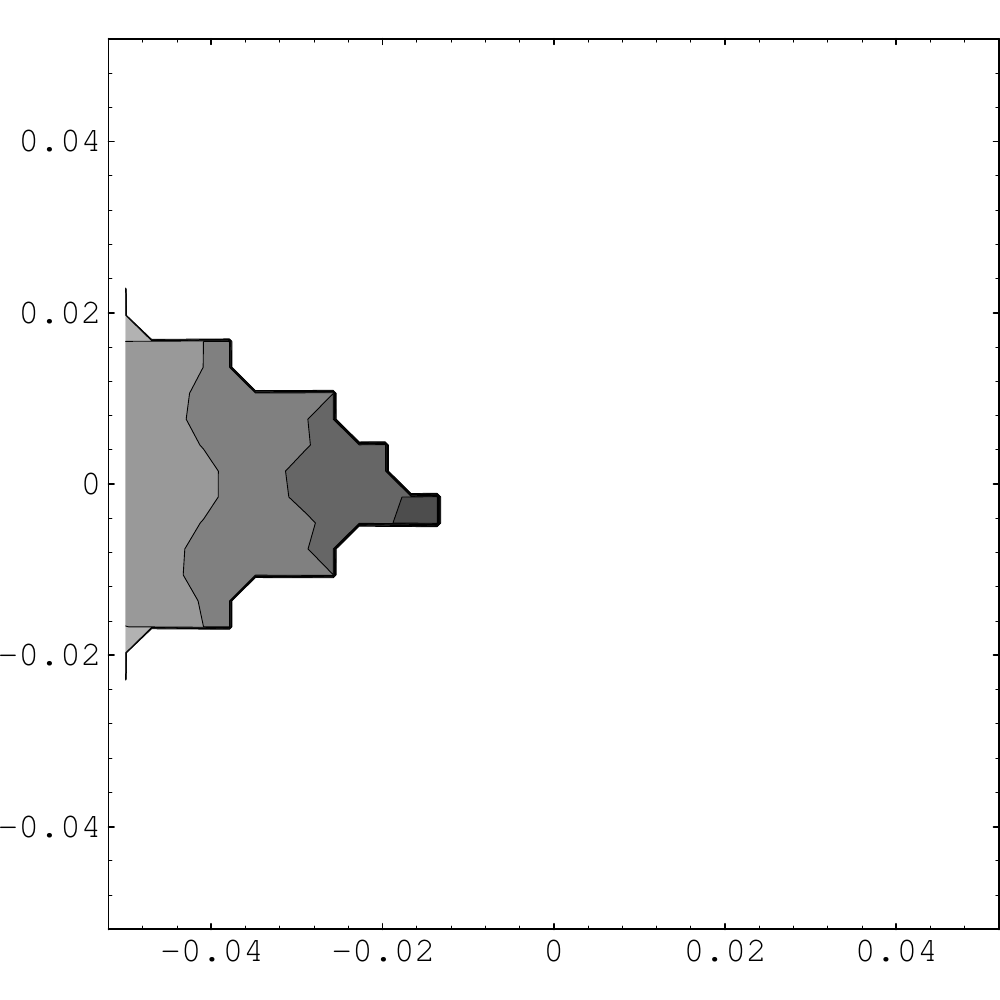}}\hspace*{\fill}\\[-0.8cm]
  \hspace*{\fill}
  \raisebox{-0.4cm}{\resizebox{0.5\linewidth}{!}{%
    \includegraphics{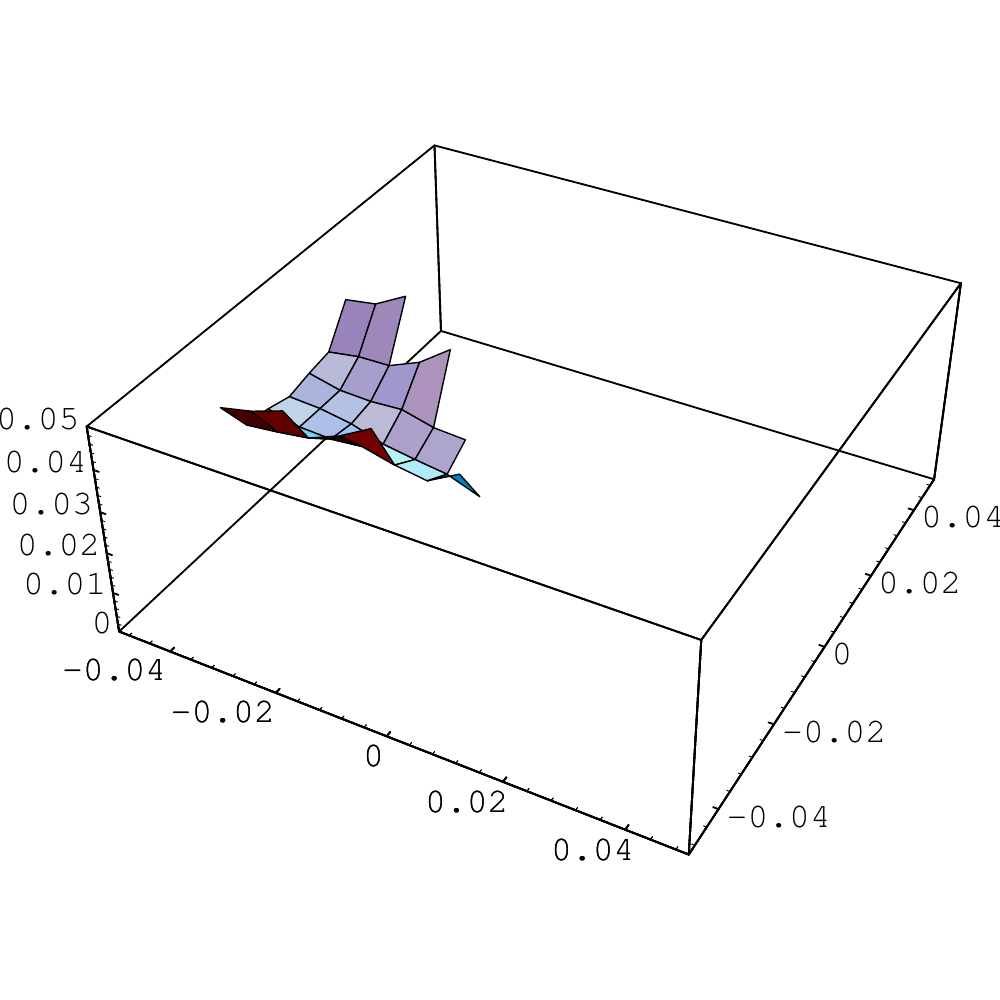}}}\
  \resizebox{0.42\linewidth}{!}{%
    \includegraphics{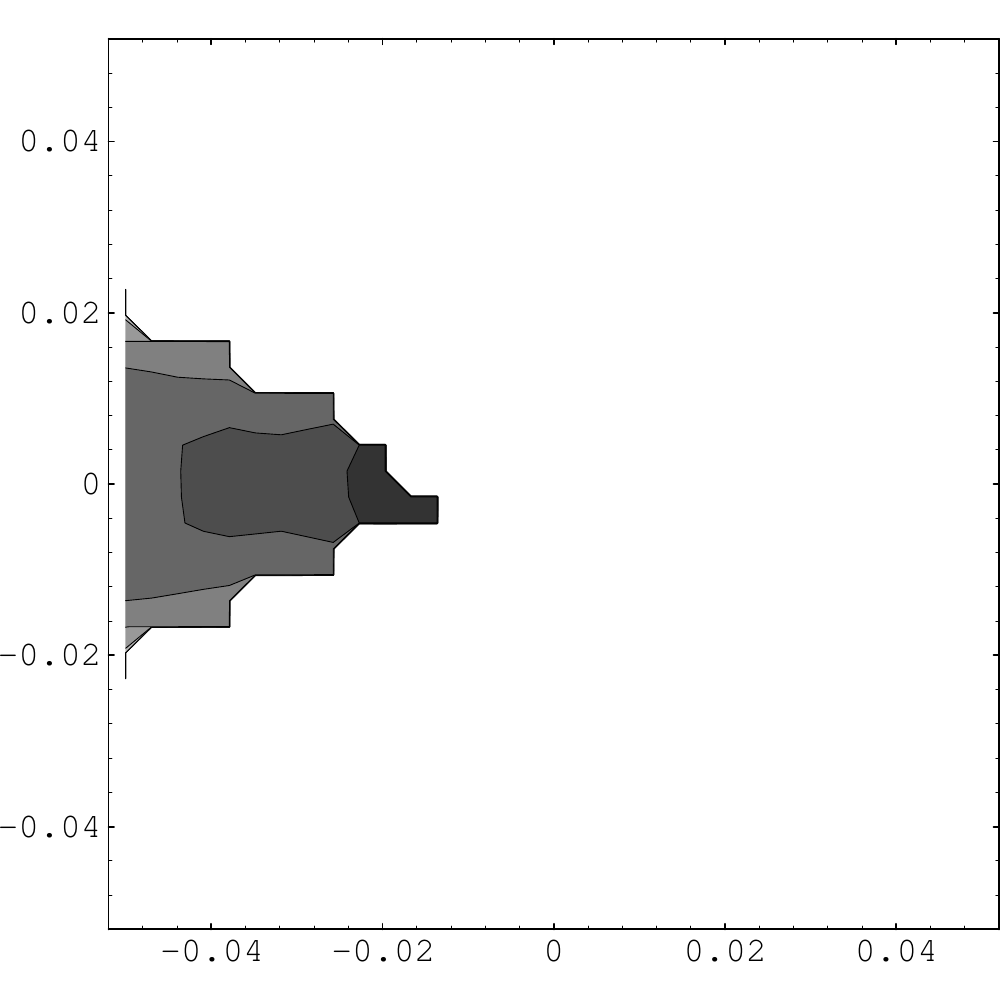}}\hspace*{\fill}%
  \vspace{-0.5cm} 
  \caption{\label{fig:bs} \small Behaviour of Bernoulli shift based systems
     in presence of S/H errors ($x$-axis: slope error $\Delta\mu$,
     $y$-axis: offset error $\sigma_n$). Top: IS bounds error; middle:
     PDF error; bottom CDF error. In contour plots darker regions
     represent lower errors, white regions are those where the system
     diverges.} 
\end{figure}

\begin{figure}[htb]
  \vspace{-0.5cm}
  \hspace*{\fill}
  \raisebox{-0.4cm}{\resizebox{0.5\linewidth}{!}{%
    \includegraphics{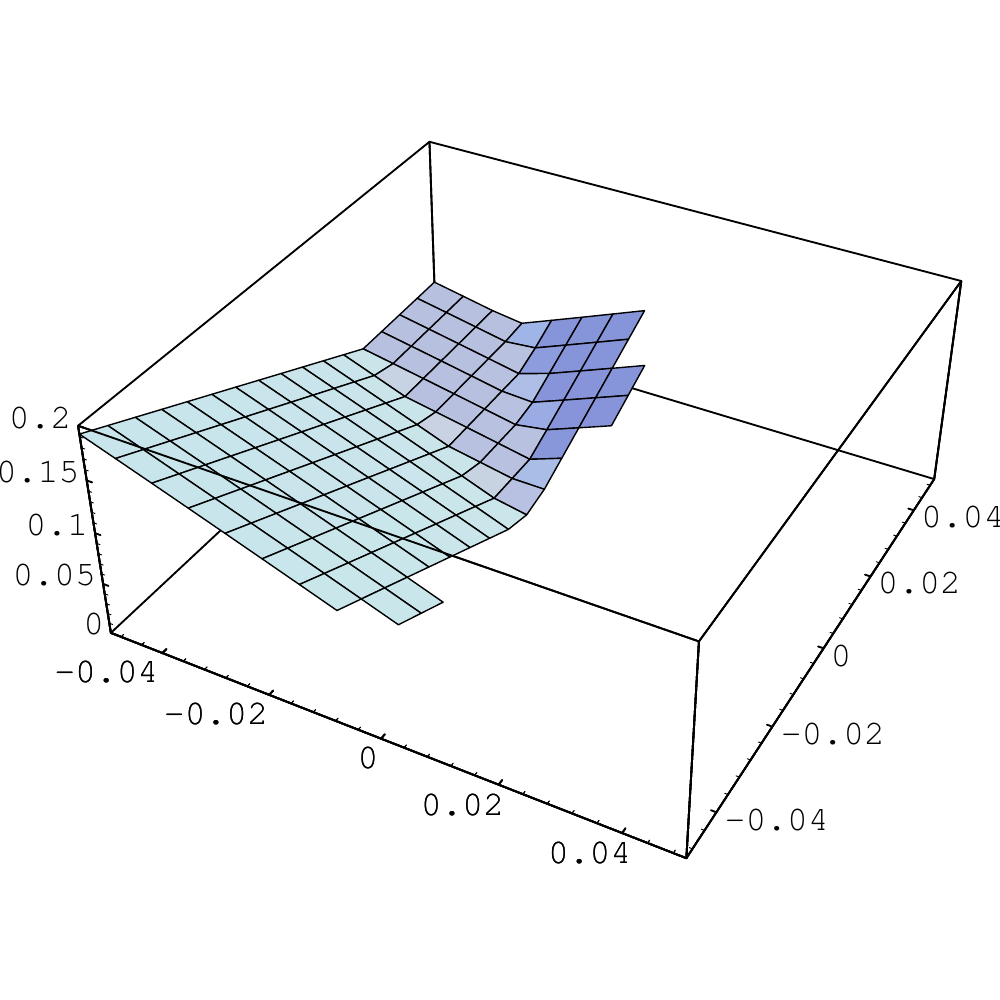}}}\
  \resizebox{0.42\linewidth}{!}{%
    \includegraphics{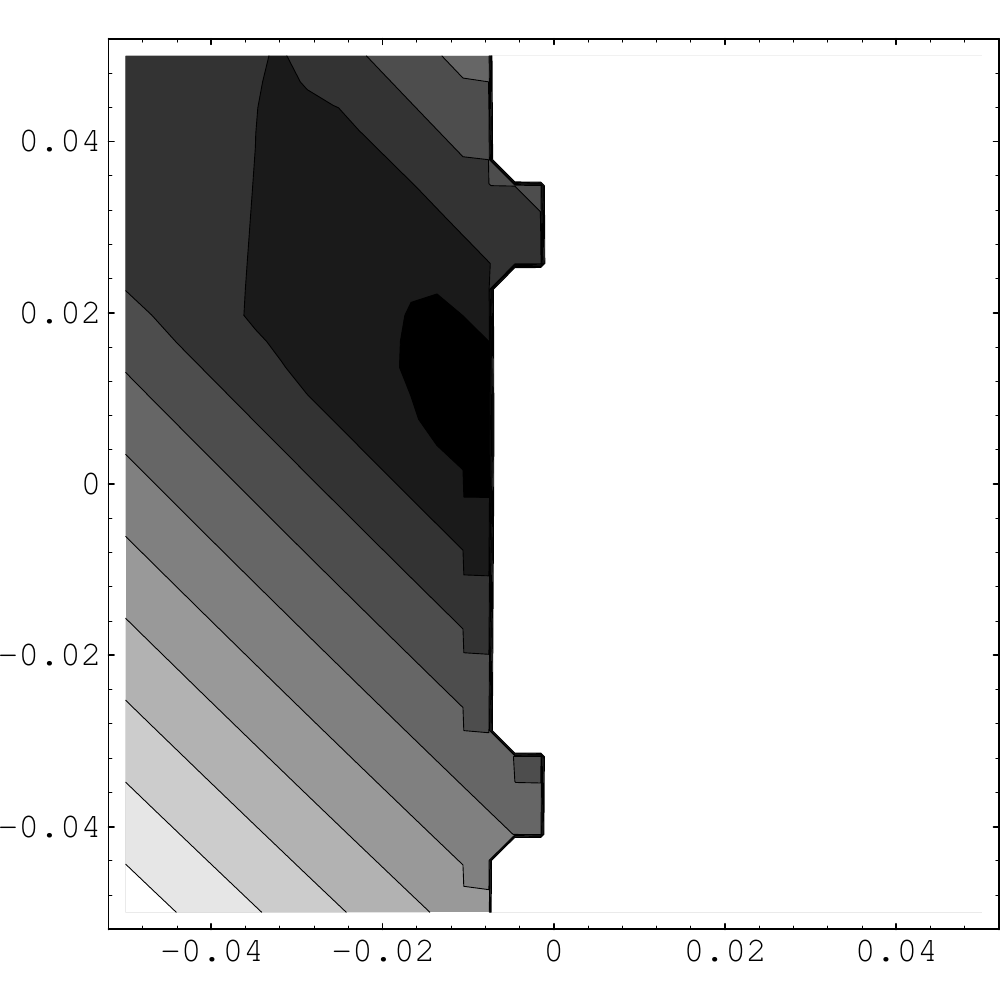}}\hspace*{\fill}\\[-0.8cm]
  \hspace*{\fill}
  \raisebox{-0.4cm}{\resizebox{0.5\linewidth}{!}{%
    \includegraphics{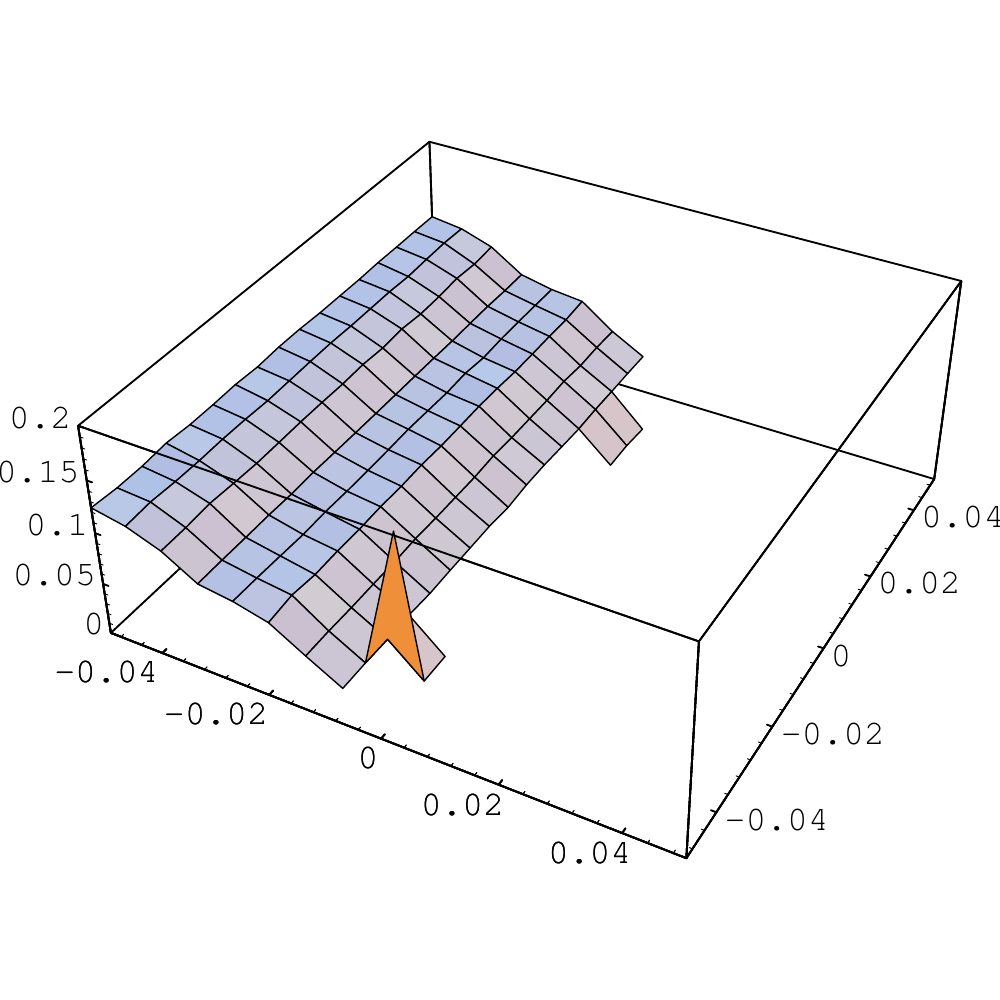}}}\
  \resizebox{0.42\linewidth}{!}{%
    \includegraphics{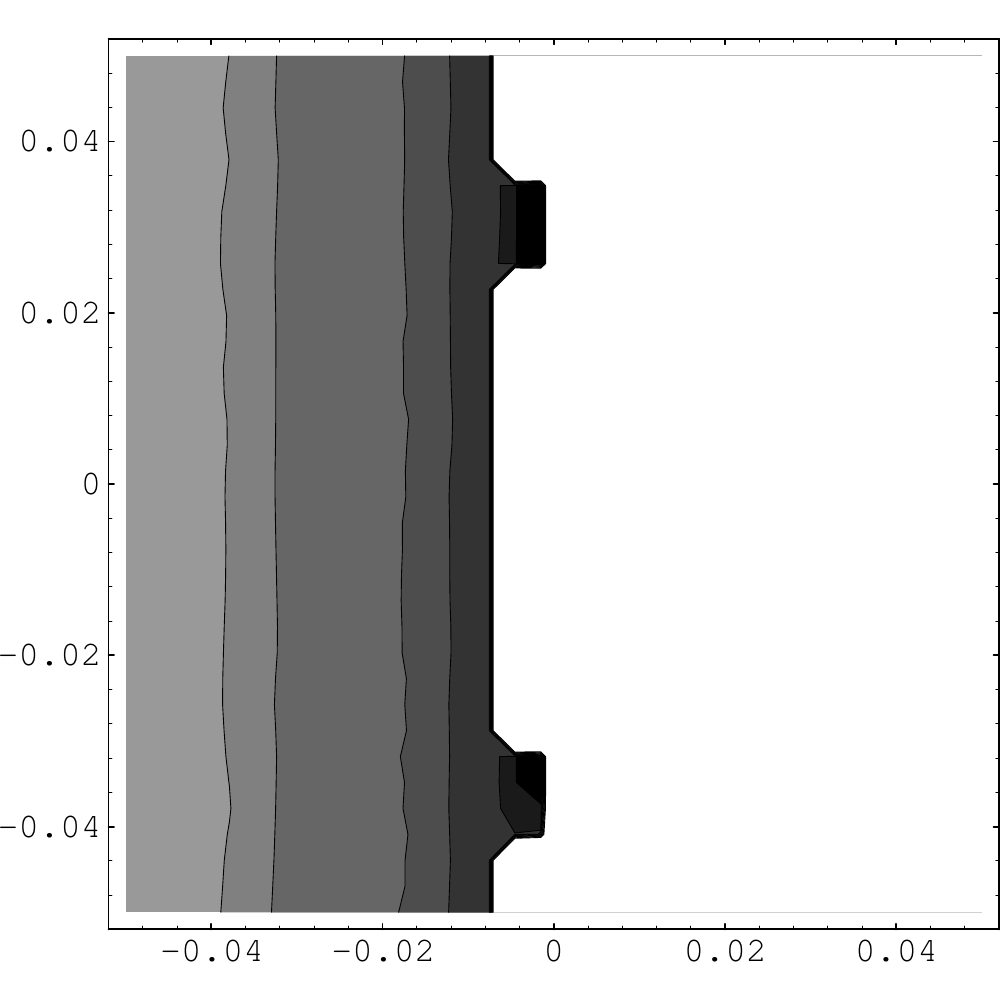}}\hspace*{\fill}\\[-0.8cm]
  \hspace*{\fill}
  \raisebox{-0.4cm}{\resizebox{0.5\linewidth}{!}{%
    \includegraphics{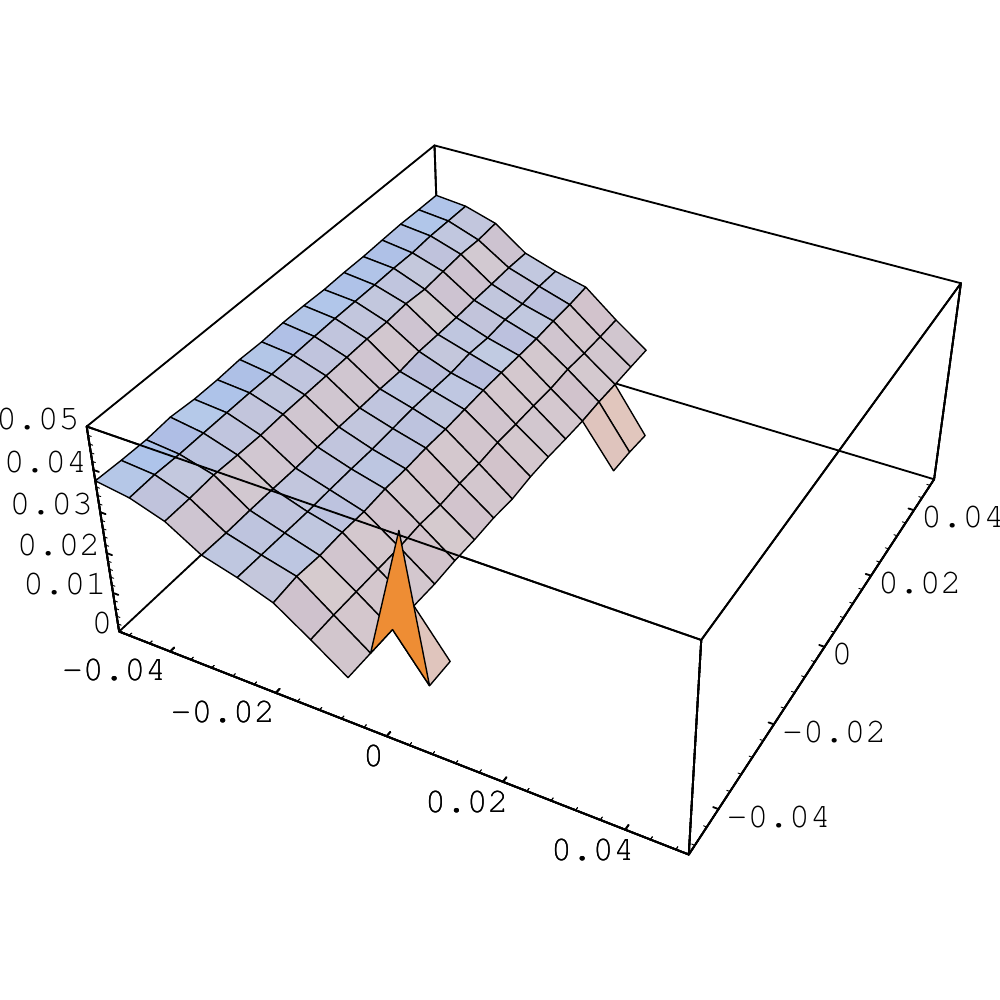}}}\
  \resizebox{0.42\linewidth}{!}{%
    \includegraphics{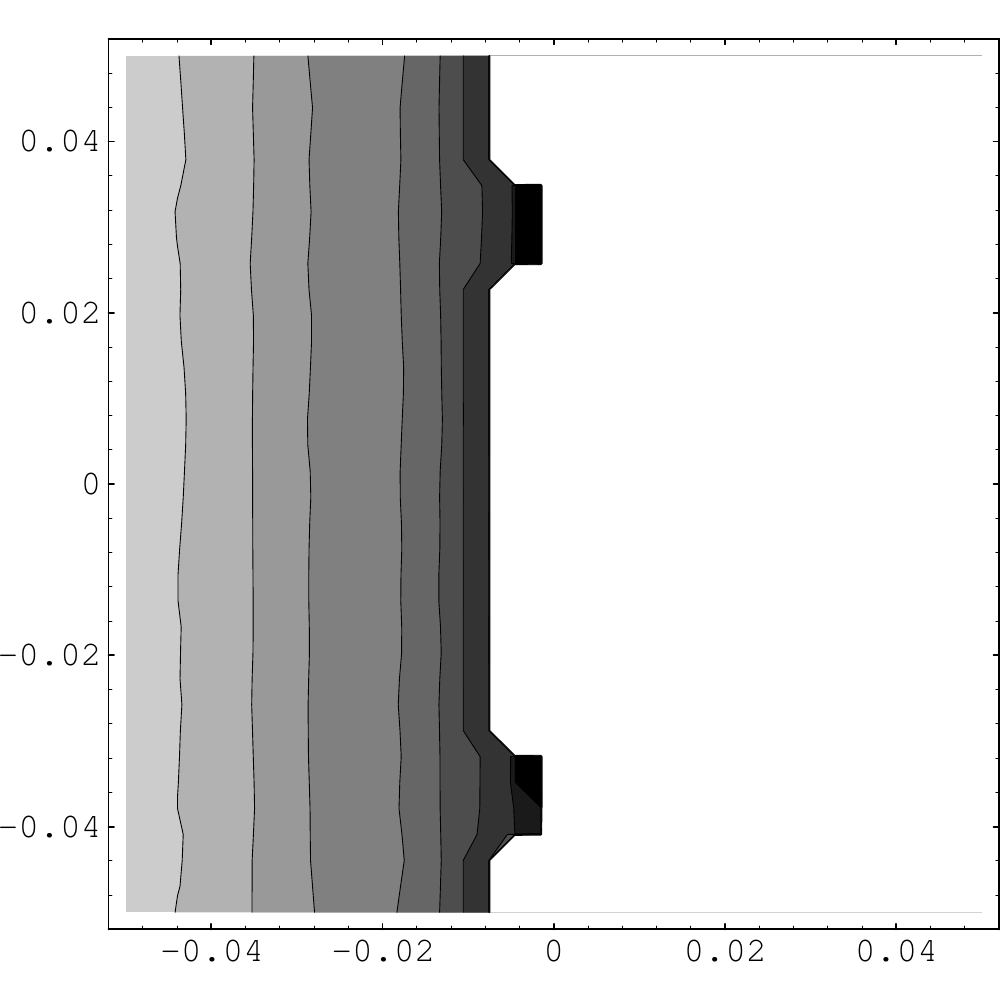}}\hspace*{\fill}%
  \vspace{-0.5cm} 
  \caption{\label{fig:tm} \small Behaviour of tent map based systems
     in presence of S/H errors. See also caption of figure \ref{fig:bs}} 
\end{figure}

\begin{figure}[htb]
  \vspace{-0.5cm}
  \hspace*{\fill}
  \raisebox{-0.4cm}{\resizebox{0.5\linewidth}{!}{%
    \includegraphics{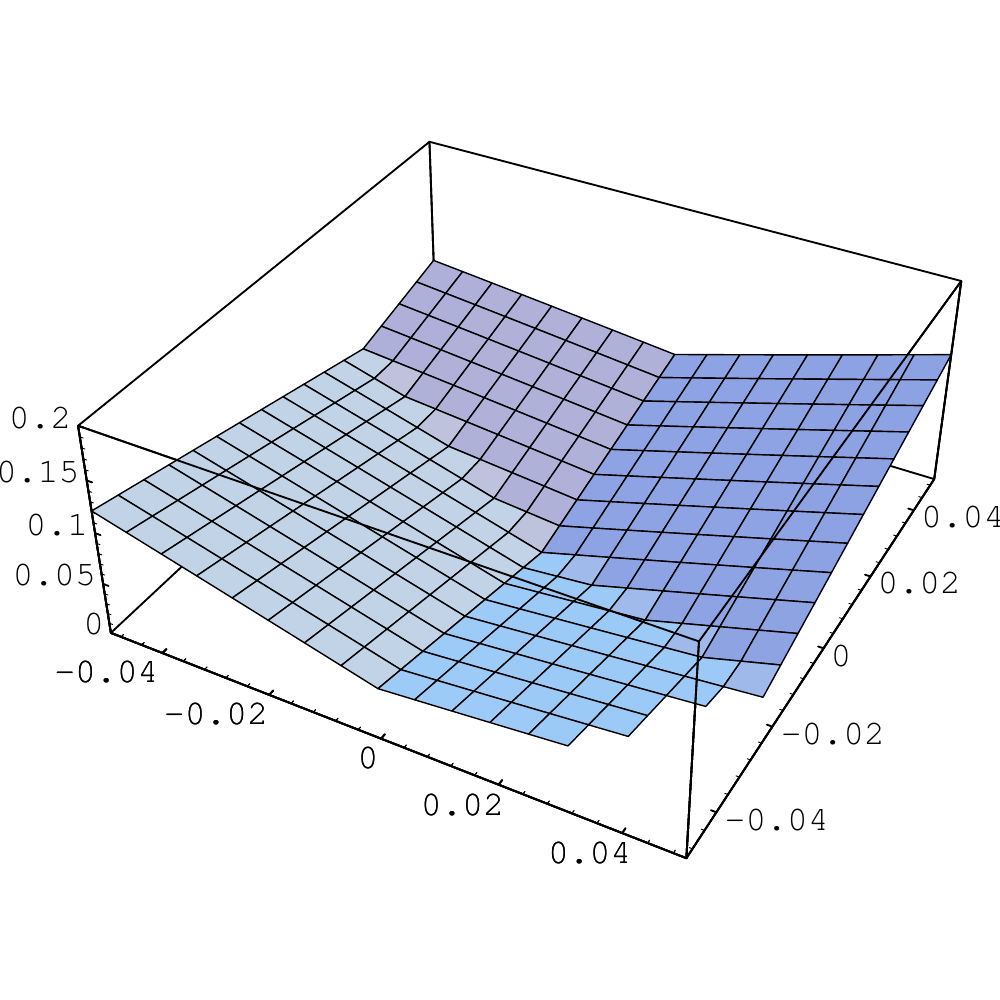}}}\
  \resizebox{0.42\linewidth}{!}{%
    \includegraphics{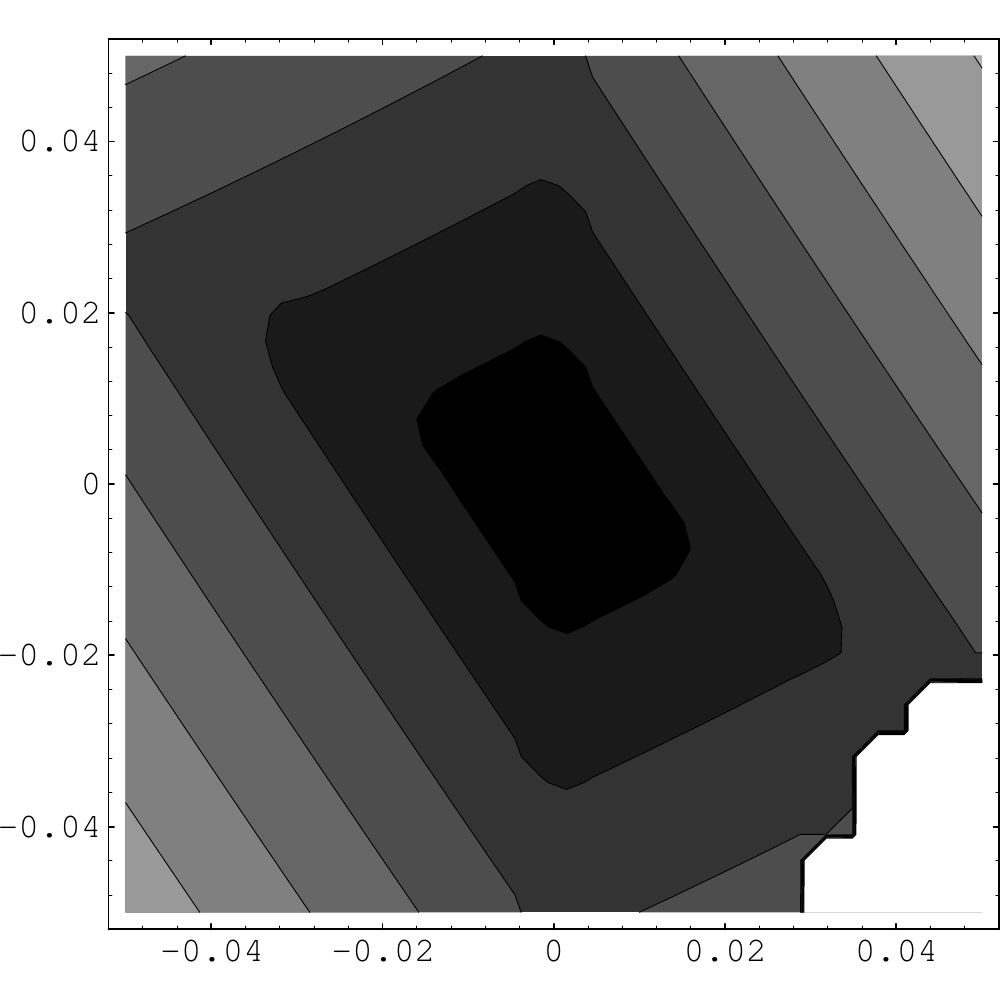}}\hspace*{\fill}\\[-0.8cm]
  \hspace*{\fill}
  \raisebox{-0.4cm}{\resizebox{0.5\linewidth}{!}{%
    \includegraphics{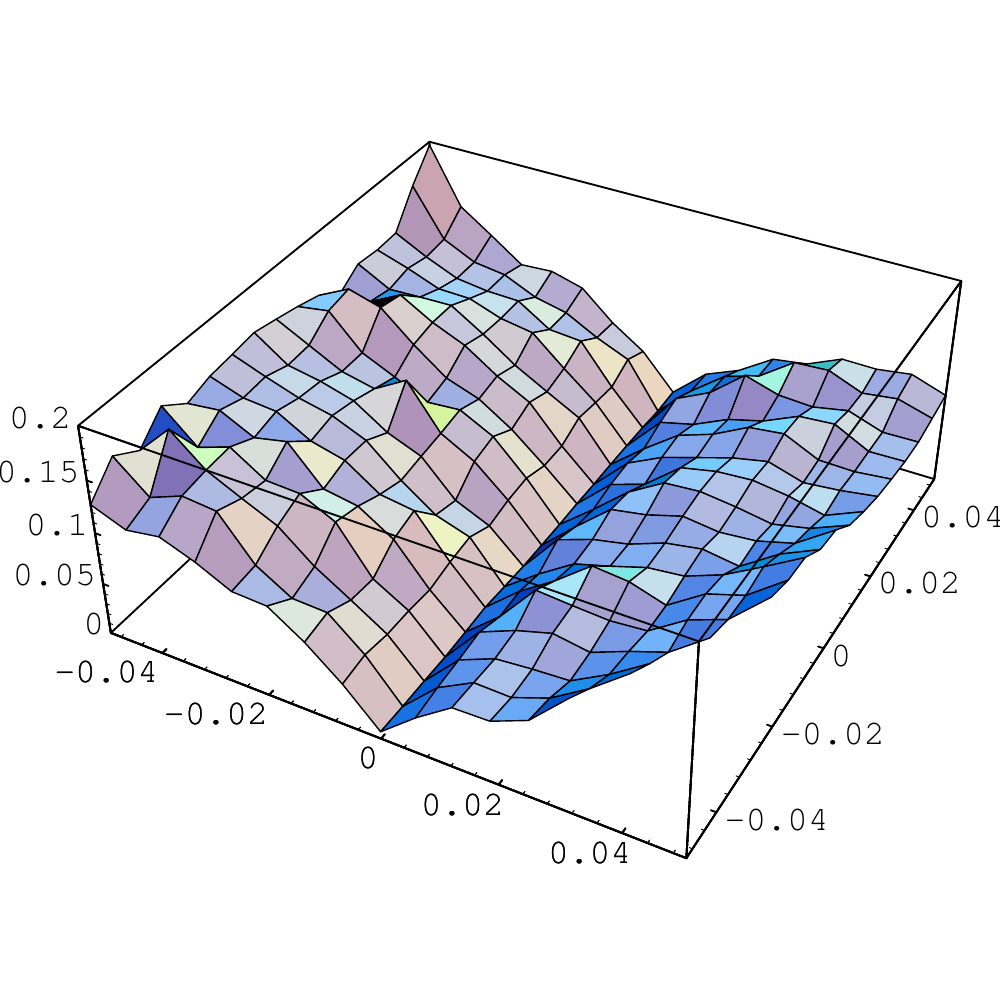}}}\
  \resizebox{0.42\linewidth}{!}{%
    \includegraphics{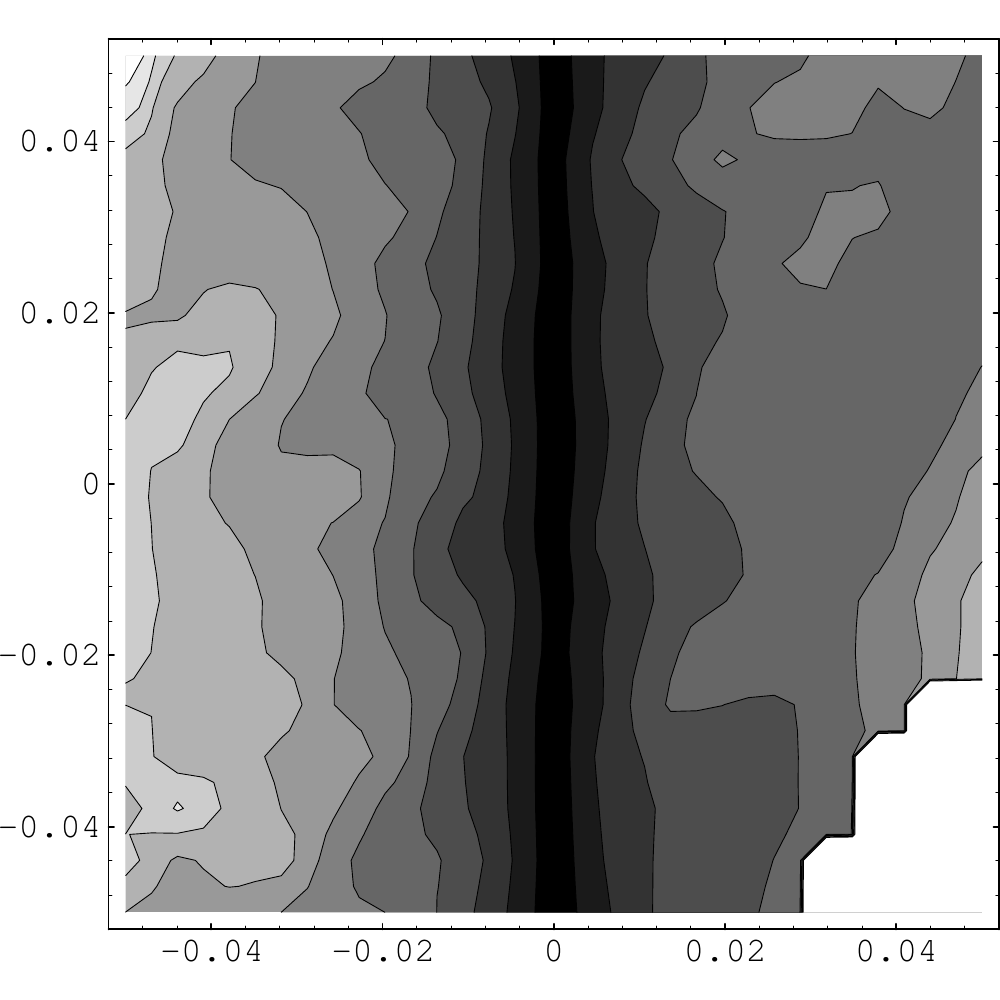}}\hspace*{\fill}\\[-0.8cm]
  \hspace*{\fill}
  \raisebox{-0.4cm}{\resizebox{0.5\linewidth}{!}{%
    \includegraphics{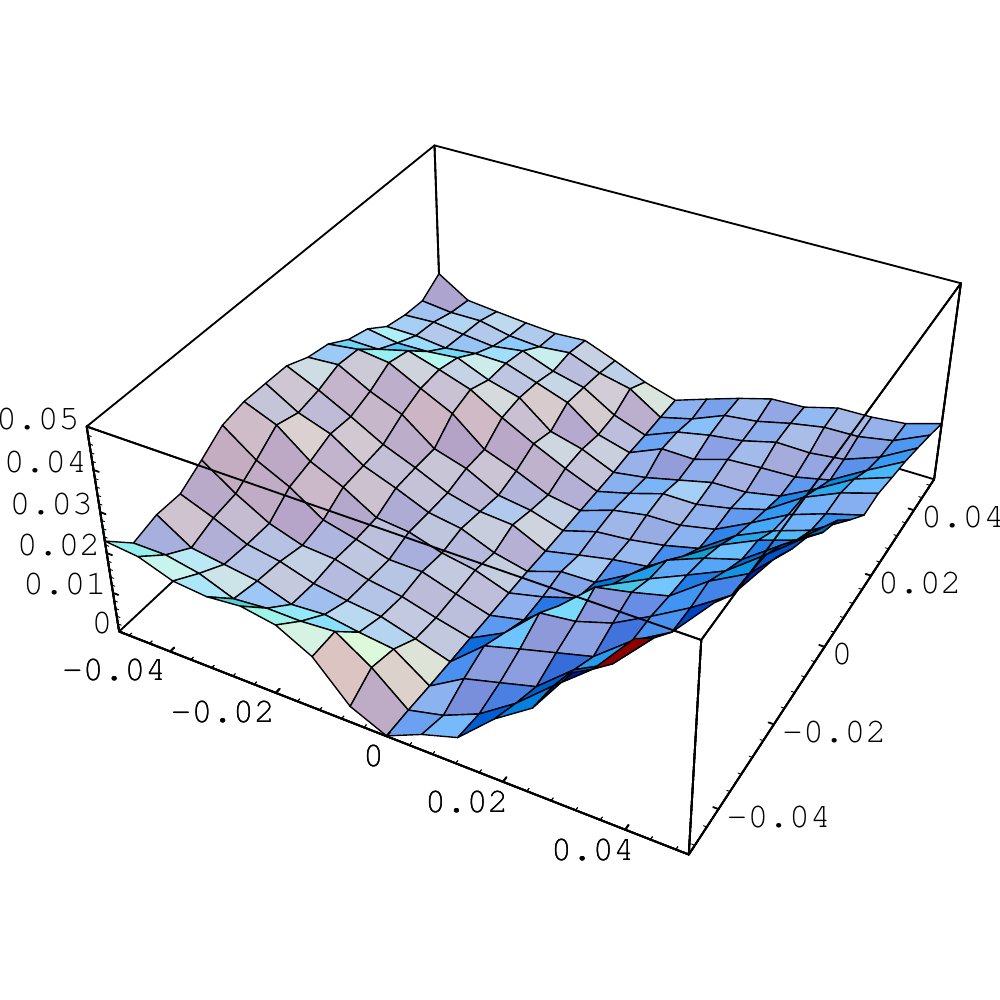}}}\
  \resizebox{0.42\linewidth}{!}{%
    \includegraphics{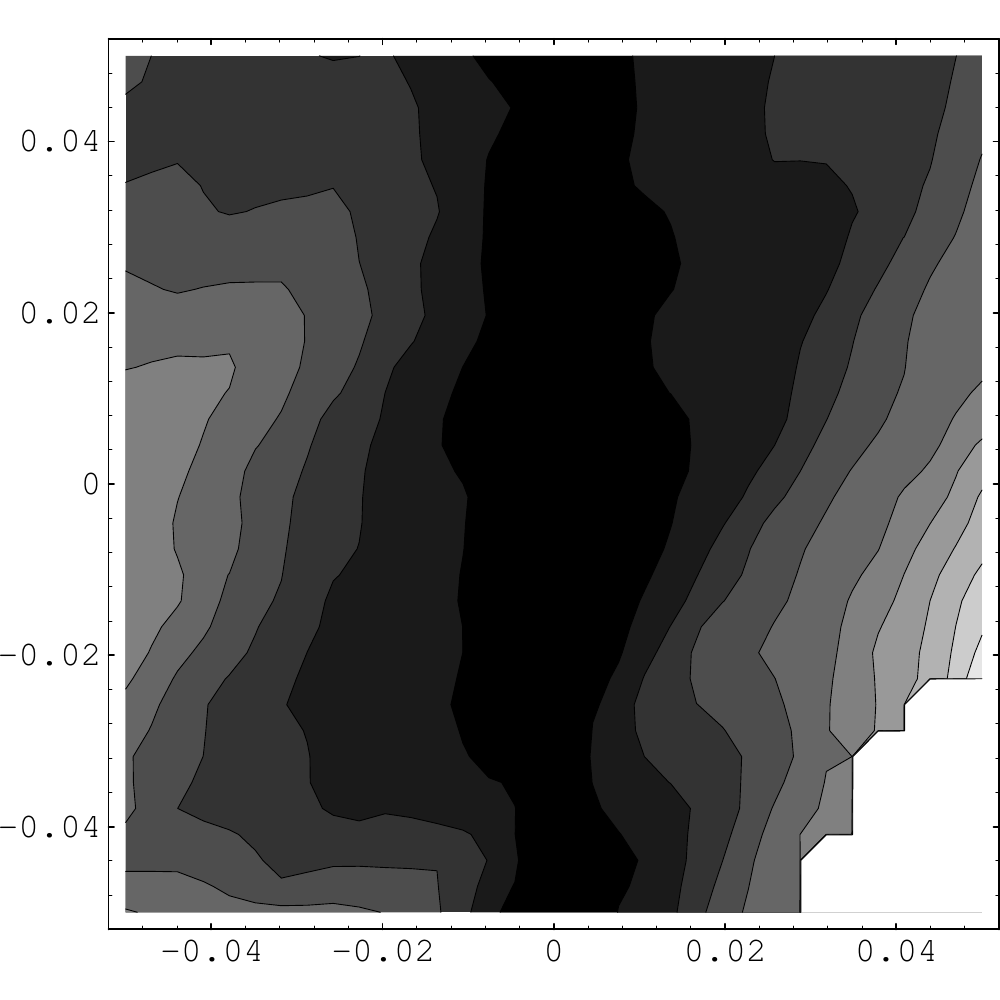}}\hspace*{\fill}%
  \vspace{-0.5cm} 
  \caption{\label{fig:ttm} \small Behaviour of tailed tent map based systems
     in presence of S/H errors. See also caption of figure \ref{fig:bs}} 
\end{figure}

Characterization of map implementations achieved by simulation is
shown in figures \ref{fig:bs}, \ref{fig:tm} and \ref{fig:ttm} (error
for a BS, a TM, and a TTM ($t=10\%$)
based system): three surfaces and contours are plotted for each map,
illustrating $\epsilon_b$, $\epsilon_\psi$ and $\epsilon_\Psi$ as a
function of $\Delta\mu$ and $\sigma_n$. White areas in the contour
plots represent loss of the system characteristic behaviour.

For what concerns robustness and alterations in the invariant set,
analytical results have been obtained in perfect accordance with
simulations, while evaluation of errors on truly statistical
properties (PDF, CDF) relies on numerical computations.

\vspace{-0.5cm}
\paragraph{Robustness.}
\label{sec:robustness}
Non-allowable values of $\Delta\mu$ and $\sigma_n$ for which systems
based on the BS, TM and TTM diverge can be determined
analytically. Namely:
\begin{equation}
  \begin{cases}
    \sh_n(0)>x_0 & \text{for}\ x_0=\sh_n(2x_0)\\
    \sh_n(1)<x_1 & \text{for}\ x_1=\sh_n(2x_1-1)
  \end{cases}
\end{equation}
is the allowable region for the Bernoulli shift, 
\begin{equation}
  \sh_n(M_n(\sh_n(1)))>x_0\quad \text{for}\ x_0=\sh_n(2x_0)\
\end{equation}
for the tent map, and
\begin{equation}
  \sh_n(M_n(\sh_n(1)))>x_0\quad \text{for}\ x_0=\sh_n(2x_0+t)\
\end{equation}
for the tailed tent map. By substituting the proposed S/H model one
gets:
\begin{gather}
  \Delta\mu<2\sigma_n\quad\wedge\quad
  \Delta\mu<-2\sigma_n\\
  \Delta\mu<0\\
  \sigma_n>\frac{2\Delta\mu^2-3\Delta\mu-2t}{2(1-2\Delta\mu)}
\end{gather}
respectively, which all find confirmation in the graphs.

In spite of the approximations in the S/H modelling, a clear tendency
emerges: perturbations increasing the steepness of the BS and the TM
($\Delta\mu>0$) lead to systems diverging to infinity. For these maps,
even ideal implementations can diverge if noise gets superimposed to
the status variable (in fact $\Delta\mu=0$ does not belong to the
allowable region). On the contrary, the TTM is tolerant to
perturbation, showing a tolerance level which is controllable via
parameter $t$ \cite{Callegari:ISCAS-1997}.

TM and BS systems can be made tolerant to perturbation by altering the
maps outside their ideal IS. For instance, a \emph{hooked} TM can be
defined as $M(x)=1-2|x-1/2|+(2+a)\,\max(-x,0)$, with $a>2$. However,
this is expensive, requiring an additional comparator to provide the
\emph{hook} breakpoint. Alternatively, robustness can be guaranteed by
designing the systems so that the map steepness is made slightly lower
than its stated value, but in this case robustness is paid in terms of
PDF and CDF accuracy. Note that a way to lower the map steepness is to
adopt a S/H topology characterized by $\Delta\mu>0$. Since this is not
difficult to be done (see next section), in some sense S/H errors can
be exploited to enhance the implementation robustness of certain
systems.

\vspace{-0.5cm}
\paragraph{Alteration of the invariant set.}
For the BS one gets:
\begin{equation}
  \label{eq:eb-bs}
  \tilde{x}_h=\sh(x_h),\quad\tilde{x}_l=\sh(x_l) 
\end{equation}
while for both the TM and the TTM
\begin{equation}
  \label{eq:eb-tm,ttm}
  \tilde{x}_h=\sh(x_h),\quad\tilde{x}_l=\sh(M(\tilde{x}_h)) 
\end{equation}
Equations \eqref{eq:eb-bs} and \eqref{eq:eb-tm,ttm} (which are valid
only for \emph{small} errors) lead to
\begin{equation}
  \epsilon_b=\left|\frac{1}{2}\Delta\mu+\sigma_n\right|+
     \left|-\frac{1}{2}\Delta\mu+\sigma_n\right|
\end{equation}
\begin{equation}
  \epsilon_b\approx
  \left|\frac{1}{2}\Delta\mu+\sigma_n\right|+
  \left|-\frac{3}{2}\Delta\mu-\sigma_n\right|
\end{equation}
\begin{equation}
  \epsilon_b\approx
  \left|\frac{1}{2}\Delta\mu+\sigma_n\right|+
  \left|-\Delta\mu\right|
\end{equation}
for the BS, TM and TTM respectively (2nd order terms are neglected).
Both the equations and the graphs, show that the TM is outperformed
both by the BS and the TTM, the latter being generally the best.

\vspace{-0.5cm}
\paragraph{Alteration of the PDF and the CDF.}
\label{sec:pdf}
As mentioned above, the analytical approach is not practicable to
evaluate $\epsilon_\psi$ and $\epsilon_\Psi$. Nonetheless, simulation
shows clearly that the three systems have similar performance in terms
of PDF, while in terms of CDF the TTM performs better.  Intuitively, a
justification comes from considering that perturbed TTM systems tend
to produce highly oscillatory PDF\emph{s} while perturbed TM systems
tend to produce rather monotonic ones. The averaging effect of the
integration needed to go from the PDF to the CDF smoothes out the
oscillations rewarding the TTM.  Note that in several
applications, notably some neural models \cite{Clarkson:WNNW93}, only
the CDF matters.

Another interesting consideration is that offset errors ($\sigma_n$)
are only a minor error source, while the major one are slope errors
($\Delta\mu$). This suggests that a better modelling of S/H\emph{s}
(i.e.\@ one that takes into account 2$^{\text{nd}}$ order derivatives)
would significantly improve the estimation of CDF errors.

\section{Implementation issues}
The analysis above may map in design guidelines.

The first, obvious consideration regards the choice of the map: the
extra price one pays for the additional comparator necessary for
the TTM may well be rewarded by the better behaviour with regard to
implementation errors. 

Secondly, TM and BS are non-robust maps: even ideal implementations
may diverge in presence of noise, so that one must adopt suitable
strategies to avoid it (note that this is true not just of the BS and
the TM: any map crossing $y=x$ at the endpoints of its IS could
present the same problem). In section \ref{sec:robustness}, it has
been shown that robustness can be guaranteed by adopting S/H circuits
characterized by $\Delta\mu<0$.  Luckily enough, the sign of
$\Delta\mu$ can be determined mathematically for many S/H topologies,
however not for all of them. For instance, S/H topologies using
\emph{dummy} and \emph{complementary switches} tie the S/H error (and
hence $\Delta\mu<0$) to the phase skew of the clock signal, which
---in turn--- is usually difficult to control.

\begin{figure}[htb]
  \addtolength{\abovedisplayskip}{-0.15cm}
  \hspace*{\fill}
  \resizebox{0.60\linewidth}{!}{\includegraphics{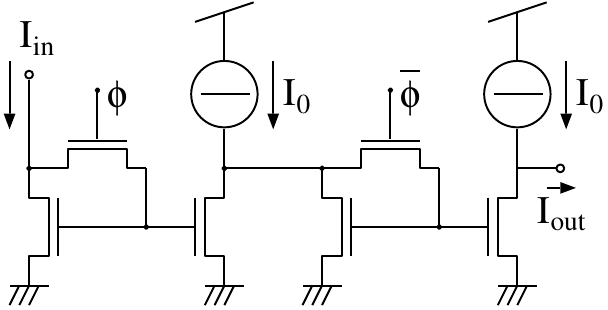}}
  \hspace*{\fill}
  \caption{\label{fig:sh-polarity}
    \small The cascade of two S/H stages aims at
    signal independent error compensation, however signal
    dependent errors still add up.}
\end{figure}

Finally, note that the most naive S/H error compensation technique
consists in cascading non-complementary stages (figure
\ref{fig:sh-polarity}), so that signal independent errors cancel each
other. While this technique could easily apply to chaotic maps, it
would lead to modest results, since signal independent errors are only
a minor error source. On the contrary, signal dependent errors ---the major
error source--- add up.

\begin{figure}[htb]
  \addtolength{\abovedisplayskip}{-0.15cm}
  \hspace*{\fill}
  \resizebox{0.56\linewidth}{!}{\includegraphics{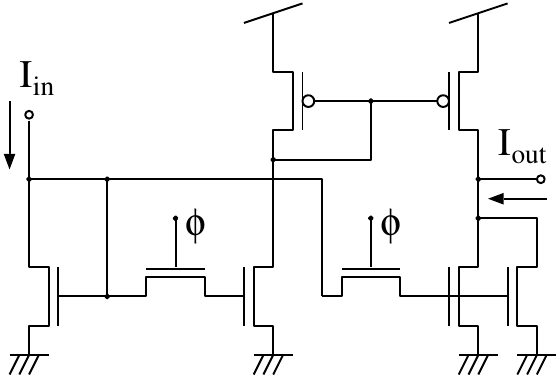}}
  \hspace*{\fill}
    \caption{\label{fig:differential}
      \small Differential S/H topologies can achieve signal dependent
      error compensation. For better performance left-hand switch
      should be twice as long as right-hand one and same width.}
\end{figure}

The correct approach to improve the implementation of chaotic maps is
to adopt S/H topologies which reduce signal dependent S/H errors. The
results presented in this contribution suggest that whenever the speed
bottleneck is given by the S/H stages due to the
accuracy/sample-latency tradeoff, the operating frequency limit might
be pushed forward by adopting such S/H topologies. Examples are
offered by differential circuits \cite{Fiez:JSSC-26-3}, figure
\ref{fig:differential} or by the S$^2$I approach
\cite{Hughes:ISCAS96}.

\bibliographystyle{ieeetr}
\small\bibliography{macros,chaos_circuits,chaos_theory,tent_map,analog,pram}

\end{document}